# Measuring Belief Dynamics on Twitter

**Joshua Introne**

School of Information Studies, Syracuse University

jeintron@syr.edu

**Abstract**

There is growing concern about misinformation and the role online media plays in social polarization. Analyzing belief dynamics is one way to enhance our understanding of these problems. Existing analytical tools, such as survey research or stance detection, lack the power to correlate contextual factors with population-level changes in belief dynamics. In this exploratory study, I present the Belief Landscape Framework, which uses data about people's professed beliefs in an online setting to measure belief dynamics with more temporal granularity than previous methods. I apply the approach to conversations about climate change on Twitter and provide initial validation by comparing the method's output to a set of hypotheses drawn from the literature on dynamic systems. My analysis indicates that the method is relatively robust to different parameter settings, and results suggest that 1) there are many stable configurations of belief on the polarizing issue of climate change and 2) that people move in predictable ways around these points. The method paves the way for more powerful tools that can be used to understand how the modern digital media ecosystem impacts collective belief dynamics and what role misinformation plays in that process.

## Introduction

How beliefs develop, change, and spread within populations is a long-standing object of scientific study with broad interest across research disciplines (Galesic et al. 2021). In an information era characterized by rampant misinformation, increasing polarization, and the rejection of conventional epistemologies, this study of belief dynamics takes on new urgency. Much work in the area has adopted sociophysics approaches, but this remains largely theoretical. There is also a well-established thread of empirical work in political science, based on longitudinal data from opinion surveys such as the American National Election Studies[1] (ANES) (e.g., Kinder and Kalmoe 2017). The approach has yielded important findings about Americans' political attitudes but has been limited to the analysis of political beliefs sampled at long time intervals and omits network-level and other kinds of variables (e.g., media exposure) that might play an important role in how beliefs change.

Social platforms allow us to bridge this gap. People use social platforms to profess their beliefs about many concerns beyond politics, including religion, science, financial markets, celebrities, consumer goods, and many other aspects of life. These beliefs can be observed longitudinally at high resolution and across a range of sociotechnical contexts. These observations will provide important additions to research on belief dynamics, and lead to new theories about how sociotechnical contexts influence belief dynamics.

With this paper, I introduce the Belief Landscape Framework (BLF) and use conversations about climate change on Twitter to demonstrate the method. I use this domain because it is highly polarized, evidenced by retweet networks. These retweet networks help both to construct a stance-aware language model, and to validate my measurement of belief clusters. The BLF yields a map of the *belief landscape*, identifying densely and lightly populated regions. I characterize this landscape and people's movements across it as a dynamic system and refer to regions of the landscape where people tend to cluster around *belief attractors*.

The contributions in this paper are primarily methodological, and I motivate the work and validate the method by examining hypotheses drawn from existing literature on belief and dynamic systems, as follows. From existing theory, some belief states are said to be more *coherent*[2] than others, and this differential coherence should hypothetically impact belief dynamics in the following ways: 1) people should remain within one or few coherent belief states over time; 2) the likelihood a person will change their belief state is correlated with how far they are from their nearest coherent state and inversely correlated with the strength of that state; 3) if they do change, they are likely to move towards states that are similar to their current state; and 4) people who have the same belief state will have similar affiliation networks.

---

[1] https://electionstudies.org/

[2] Unlike prior work (Converse 2006), I do not use this term to imply normative assumptions; instead, coherence is an empirical phenomenon.

The term "hypothesis" should not signal the reader that this paper adopts the hypothetico-deductive method (Godfrey-Smith 2021) used in science to evaluate an established theory. Instead, I use hypotheses that are logical extensions of existing theory to set expectations about what kinds of findings a well-functioning method should produce given current theories of belief dynamics. I do this because there is little existing data that might be used for validation. I investigate these hypotheses across a range of parameter settings to further build confidence in the method.

This paper shows that the BLF yields data that conform to these hypotheses, and its performance is robust across a range of parameter settings. I also provide a qualitative analysis that shows the belief landscape exhibits a plausible spatial arrangement of beliefs on Twitter, capturing nuances that are not normally considered in stance detection literature.

## Literature

The study of belief dynamics is expansive, having been examined across a broad range of disciplines, including biology, psychology, anthropology, sociology, and statistical physics. Despite this attention, beliefs are extremely hard to study in the wild. In the abstract, a belief is a truth value or degree of confidence that a person assigns to a proposition. What it means for an individual to hold a given belief is less easily stated; people's beliefs can vary over time, from context to context, or even conflict with one another. For instance, research has found that some who believe in conspiracy theories maintain logically inconsistent beliefs at the same time (e.g., Princess Diana was assassinated and staged her own death) (Wood, Douglas, and Sutton 2012). Such inconsistencies are not restricted to conspiracy theorists, leading some to propose that people maintain modular belief systems that are active in different contexts (Porot and Mandelbaum 2021). Given this inherent inscrutability, I do not propose to unpack the authenticity of a stated belief. A belief uttered on social media is as valid a datapoint as one that is assessed in a survey on one's political attitudes.

Inconsistencies aside, an important aspect of belief is that beliefs are positively or negatively correlated with one another, suggesting that a set of constraints govern their interactions. These relationships play a central role in both empirical and more recent theoretical work. In his seminal work, Converse (2006) described the set of constraints as a *belief system*. In his analysis, the conventional Western conservative-liberal spectrum is a single belief system wherein "coherent" subsets of beliefs lie at either end of the spectrum. A maximally coherent set of held beliefs is one that violates the least number of constraints within a presumed belief system. This model is closely related to Thagard's psychological model of attitude coherence (Thagard 2002).

Converse was vague about what governs the constraints amongst beliefs—it may be logical entailment (c.f., Friedkin and Bullo 2017) but might also stem from cultural associations or social pressure. Nonetheless, in his empirical research, Converse presumed the existence of a single, ideal political belief system. Analyzing survey data collected as part of the ANES project, he found that political sophistication correlates with belief coherence (i.e., more uniformly conservative or liberal beliefs). This result has held in many analyses since Converse's time (Kinder and Kalmoe 2017).

Another explanation for Converse's findings is that there is greater diversity among belief systems than Converse and others have presumed (Baldassarri and Goldberg 2014). It is not unreasonable to imagine multiple coherent belief configurations, leading to renewed attempts to infer constraints from survey data (Boutyline and Vaisey 2017). Nonetheless, the constraint structure of belief systems is not critical to studying belief dynamics if pockets of coherent beliefs and people's movements among them could be assessed directly. Such an analysis would enable researchers to examine, among other things, causal relationships between media ecosystems (e.g., misinformation, media bias, algorithmic filters, etc.) and belief dynamics.

Social media is an obvious source for this kind of data. Whereas few efforts have sought to examine how a broad spectrum of different beliefs co-occur on social media, substantial research has been devoted to inferring peoples' stances with respect to a single belief. For instance, one challenge in the 2016 SemEval workshop focused on stance detection using Twitter data (Mohammad et al. 2016). Stance detection involves detecting an individual's disposition with respect to a specific target (e.g., whether climate change is a real phenomenon). The workshop posed both supervised and unsupervised challenges. I focus on approaches to the unsupervised task here because this is more closely related to the task of extracting beliefs in general.

SemEval 2016 preceded the development of modern transformer-based language models (Devlin et al. 2019), and the best team achieved an average F-score of 56.28 using a deep learning approach (Wei et al. 2016). Since that time, the area has continued to receive increasing attention (ALDayel and Magdy 2021), and performance has increased steadily. One important innovation has been the use of social network data. Social networks had been previously shown to be of value in estimating the ideological position of individuals (Barberá 2015). Building on this idea, Darwish et al. (2020) showed that clustering users according to their retweeted accounts (an affiliation network), could reliably identify users with different stances. More recently Rashed et al. (2021) introduced a related method for unsupervised stance detection. Instead of using retweeted accounts, they used sentence embeddings derived from Google's Multilingual Universal Sentence Encoder (Y. Yang et al. 2019), and averaged sentence embeddings across a filtered set of tweets for each user. This approach achieved slightly lower precision but higher recall than the method introduced in Darwish et al. (2020). Notably, both methods produced more granular stances than a simple "for" or "against" a set of target propositions.

Here, I combine aspects of both Darwish et al. (2020) and Rashed et al. (2021). Instead of using retweet clusters directly, I use them to train a language model and use this language model to extract beliefs across a range of topics. Because I investigate how beliefs co-occur, the BLF identifies more nuanced beliefs than simple bivalent stances. Absent gold data for stablishing accuracy, I offer partial validation by comparing my results to inferred user stances as well as a set of predictions from prior work in opinion dynamics and complex systems. I discuss some of this work here.

### Belief Dynamics

Many studies of belief dynamics occur in the sub-field of opinion dynamics, which often considers the conditions under which individuals will achieve consensus on an issue. One influential model is DeGroot's model of consensus formation (Degroot 1974), which proposes a social learning process in which individuals update their own beliefs according to the average of their current beliefs and those of their associations, subject to the trust they place in these associations. Whether beliefs converge to consensus depend entirely on the structure of the trust matrix.

Like many models of social contagion, DeGroot's model does not incorporate a model of constraints between beliefs (or contagions more generally), and several models have sought to address these limitations. Lakkaraju and Speed (2019) explored an agent-based model in which agents both exchanged beliefs but also sought to maintain cognitive consistency. Friedkin (Friedkin and Bullo 2017; Friedkin et al. 2016) extended DeGroot's model to incorporate a set of constraints that capture logical relationships between beliefs. More recently, Galesic et al. (2021) present a general model that incorporates both cognitive dissonance (due to violation of constraints among beliefs) and social dissonance (similar to DeGroot's consensus model). Nonetheless, these richer models have not yet been used to investigate the broader dynamics of belief in the context of multiple coherent belief states. A deeper understanding of these dynamics would help researchers make additional headway on questions about how media and misinformation affect belief at scale.

Other work provides clues about how multiple coherent belief states might impact dynamics. Political science suggest that ideologues (those with more coherent belief structures) are less likely to change their minds because their beliefs are more highly constrained (Converse 2006; Sartori 1969), and this prediction has been demonstrated via simulation studies (Brandt and Sleegers 2021). Recent work has also provided evidence that belief interventions cascade to non-targeted beliefs (Turner-Zwinkels and Brandt 2022), suggesting that small belief changes may hasten an individual's movement from one stable belief state to another.

From this context, one can infer that people update their beliefs over time as if traversing a landscape composed of stable regions (where beliefs change slowly) and transitional regions (where beliefs change more quickly). In the language of dynamic systems, this field of belief states can be described as an *attractor landscape* (Miller and Page 2009), where each stable belief state is an attractor. The work reviewed here suggests there is a generally smooth gradient across this landscape—in other words, neighboring points (similar belief states) are more likely to have similar levels of coherence than distant points (different belief states).

With this framing in mind, I propose four hypotheses as an initial test of the BLF. I refer to a stable belief state as an attractor, and the entire space as a belief landscape.

*H1:* People will tend to visit few attractors on the belief landscape (i.e., people are unlikely to change their beliefs much over time).

*H2:* Individuals who occupy more peripheral regions of a belief attractor or are near weak attractors are more likely than others to transition to new attractors.

*H3:* Transitions between adjacent belief attractors are more likely than transitions between more distant attractors (i.e., people who do update their beliefs do so incrementally)

*H4:* People who are closer together on the belief landscape will have similar affiliation networks (i.e., people who believe the same thing tend to retweet the same alters).

As discussed, I do not seek to prove these hypotheses. They are used as an initial conceptual framework to evaluate the outcome of the BLF methodology. I discuss one approach to developing stronger tests in the discussion section.

## Data & Methods[3]

### Data Preparation

For this study, I examined five months of tweets between May 23, 2020 and October 27, 2020. Twitter's V.1 streaming API, which retrieved up to 1% of all Twitter traffic, was used to collect tweets containing terms and phrases related to climate change, including "climate," "climate change," and "global warming." The dataset includes approximately 29.3 million tweets from 5.9 million users. I focus on climate change because it remains a polarized discussion, and I used this polarization both to enhance my pipeline (discussed further below) and provide partial validation for my results. Of the 29.3 million tweets collected, 4.7 million of these were in English (from about 1.5 million unique users), and these formed the dataset for the study. I focused on English tweets for this study because the grammatical structure of English simplifies the process of extracting belief-like statements (see Step 1 in the methods section).

---
[3] Code and data available at https://github.com/c4-lab/BeliefLandscape-Framework

As discussed, I determined active user's stances regarding climate change following Darwish et al. (2020). I focused on the 25,981 users who tweeted at least 100 times and had more than one retweet, and applied UMAP (McInnes, Healy, and Melville 2020) and HDBSCAN (McInnes, Healy, and Astels 2017) to cluster these users based on the accounts they retweeted. 82.6% of the accounts examined belonged to climate believers, 16.7 % to climate skeptics, and the remaining accounts could not be clustered.

To establish label quality, I used MTurk workers to label 2733 randomly selected accounts as believers or skeptics (defined as people who did not believe in climate change or believed the climate change was not anthropogenic). Three workers labeled each account based on a sample of 5 Tweets and the account profile description, with the final label determined by the majority. Based on this data, the "climate believer" cluster had a purity of 99.9%, and the "climate skeptic" cluster had a purity of 80.7%. I examined half of the 34 users labeled as believers by MTurk workers but that clustered with skeptics and found the majority (fifteen) either did not appear to believe in anthropogenic climate change or did not believe there was anything that could or should be done. Therefore, I chose to use the labels established by the clustering procedure as the best indicator of stance.

## Method

Figure 1 provides a high-level schematic of my overall method. I identified prevalent belief propositions (the first two steps in Figure 1), collected these into representative belief *states* (spanning multiple beliefs) to establish the structure the landscape, and measured peoples' movements across it (Step 3). I discuss each of these steps below.

**Step 1: Extracting belief statements.**
To identify belief-like utterances, I extract declarative statements that could be parsed into subject-verb-object (SVO) tuples. This omits questions, (e.g., "Why do Democrats continue to lie about climate change?") and complex multi-clausal statements (e.g., "Lie, cheat, steal… that's how climate change alarmists roll!") but covers a large proportion of belief-like statements in English, which is a predominantly SVO language (Tomlin 2013).

To do this, I built a plugin for the SpaCy open-source library (Honnibal and Montani 2017). The plugin performs pattern matching on top of the dependency parse, first identifying the root verb in a clause, and then a subject and object. The plugin uses a co-reference resolution plugin (Coreferee[4]) to resolve pronominal subjects in the context of a given tweet, corrects for negated beliefs, flags attribute beliefs, and handles nested constructions.

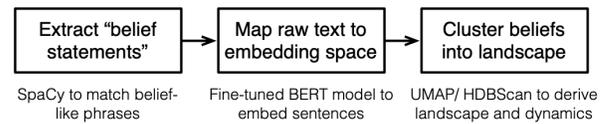

Figure 1. Overview of the data analytic pipeline

I examined the algorithm's performance using a manually curated set of 50 belief-like Tweets containing SVO tuples with varying degrees of complexity. The routine performed well, but there were several common failure modes. The parser extracts complex noun phrases that are too specific given the need to recognize commonly held beliefs. For example, given the sentence "The current ideology behind the climate change agenda is a problem." the parser extracts "The current ideology behind the Climate Change agenda" as the subject. Another problem, specific to Twitter, stems from the use of handles in sentences. For example, for the passage, "The @DEC's Justin Good @DECGood marvels: People don't like wind, don't like nuclear, but 'everyone loves hydrogen'," the parser extracts the subject "The DEC's Justin Good DECGood." Aggressive pruning of mentions would reduce this problem, but Twitter users often use handles to stand for proper names, so I retained all mentions. Finally, many tweets were rife with spelling errors and ungrammatical constructions. These often failed to parse, but sometimes generate anomalous parses.

The parser identified about 949k sentences containing belief-like statements spanning roughly 311k unique subjects (of the extracted subject-verb-object tuples). Many subjects differed by a leading article (e.g., "the climate crisis" vs. "climate crisis"), several subjects differed by an intervening space (e.g., "climate change" vs. "climatechange"), and salient individuals' names exhibited multiple permutations ("joe biden" vs. "biden" vs. "joebiden"). After addressing these issues and removing deictic references (e.g., "he" or "this")[5] 288k unique subjects remained, distributed as a power law (the top two subjects were "we" and "climate change," accounting for roughly 7% and 3% of the data, respectively). To narrow the data to representative beliefs, and observing that believers were over-represented, I ranked the subjects for believers and skeptics separately, averaged ranks across the two groups, and chose the top 100 subjects. These subjects covered 170,437 sentences, which I will refer to as the focal set of sentences.

---

[4] https://github.com/explosion/coreferee

[5] I chose to leave the deictic reference "we," because this is commonly used to refer to the human species in the context of climate conversations.

**Step 2: Mapping beliefs to an embedding space.**
The goal of Step 2 was to map the focal set of sentences into a vector space that encodes semantic relationships and then cluster the data to identify highly similar belief propositions. I adopted Sentence-BERT (Reimers and Gurevych 2019) for embedding text into a vector space.

Many pre-trained models are available in the HuggingFace transformer library (Wolf et al. 2020), and I sought to select amongst them by evaluating their ability to distinguish a small set of probe sentences (e.g., "Climate change is a serious threat," "Climate change is a liberal conspiracy"). All tested models[6] performed poorly on the probe task. For example, the S-BERT model trained using the "all-mpnet-base-v2" model (recommended for general use by the creators of S-BERT (Reimers 2023)) reports a cosine similarity between the phrases "climate change is real" and "climate change is not real" of .9.

To address this, I fine-tuned a pre-trained language model (the paraphrase-MiniLM-L6-v2 model[7], which exhibited relatively good performance on my initial probe task), using a sentence similarity task (Reimers and Gurevych 2019). To create the training set, I sampled pairs of sentences from the full set of belief-laden sentences and calculated cosine similarity using that reported by the untuned model and knowledge of user stances. My goal was to strike an effective balance between preserving semantic relationships while capturing fine-grained positional differences.

I considered two strategies for estimating the similarity of sentences in my training set. For sentences from homogeneous users (with the same stance), I used the similarity reported by the untuned model. For other (heterogeneous) pairs, I explored two different approaches; the ***min*** approach simply returns -1, and the ***invert*** approach negated positive values from the untuned model.

I applied both similarity strategies to each of three different sampling methods to create the training sets:

- ***Random***: 0.5 million homogeneous sentence pairs were selected from each stance along with 1.5 million heterogeneous sentence pairs. I used a *cutoff* similarity value to determine how similar (in the untuned model) heterogeneous sentences needed to be to apply the chosen similarity manipulation (i.e., ***invert*** or ***min***).
- ***KNN-based***: 0.5 million homogeneous pairs were selected from each stance. To select heterogeneous pairs, I selected the k-nearest neighbors for each of the "skeptic" sentences in the dataset.
- ***Cluster-based***: The untuned model was first clustered using UMAP and HDBSCAN. For clusters with a given *purity cutoff,* I randomly selected .001 of all sentence pairs in each such cluster, balanced across believer, skeptic, and heterogeneous pairs.

For each training set, I used three training epochs with the default learning rate[8]. Figure 2 compares the embedding space of the untuned model with one of the fine-tuned models (***random***, ***min***, with a *cutoff* of 0.8). The general effect of training was to separate skeptics' sentences from believers' while preserving some (though not all) of the semantic distinctions present in the untuned model.

To evaluate the different approaches, I clustered sentences using UMAP to reduce the dimensionality of the embedded data to two dimensions and applying HDBSCAN and labeled each cluster by its majority stance. I used visual inspection of the untuned model to set the parameters for these algorithms, as follows. Using the defaults for both libraries[9], data clustered in the center of the visualization but there was also a large cloud of peripheral points. Fixing the *min_dist* cutoff at (0.1) I raised the UMAP *nearest_neighbors* parameter from a minimum of 2 until data in the center of the visualization began to form visually distinct clusters (*nearest_neighbors* = 20). I then examined value settings higher than the default (0.1) for *min_dist*, but this caused clusters to become merged. To set HDBSCAN parameters (*min_samples* and *min_cluster_size*), I explored the parameter space to achieve an optimal clustering of subjects extracted from SpaCy (i.e., each cluster was predominantly populated by tweets with a single, common subject), clustering as much of the central region as possible while avoiding spurious clustering of outliers, arriving at a final setting of *min_samples*=100 and *min_cluster_size*=200.

To evaluate the different models, I calculated three measures for different stance clusters: the number of clusters, the percent of tweets that could be clustered (coverage), and the average purity of each cluster. The purity of a cluster is the percentage of clustered items that have the same class as the most representative class of items in the cluster. More precisely, given a set of possible item classes **C**, which in this case is {*skeptics*, *believers*}, and a cluster **K** of items $\{k_0, k_1, \ldots, k_n : k \in \mathbf{C}\}$, then let the label of that cluster be defined as:

---

[6] Including: all-mpnet-base-v2, all-distilroberta-v1, multi-qa-mpnet-base-dot-v1, bert-base-nli-mean-tokens, paraphrase-MiniLM-L6-v2, bertweet-covid19-base-uncased, bertweet_retrained_semEval2019, poleval2021-task4-herbert-large-encoder, distilbert-political-tweets, stsb-bertweet-base-v0, bertweet-base-stance-climate, climate-fever-msmarco-distilbert-gpl, distilbert-similarity-b32, paraphrase-mpnet-base-v2_SetFit_sst2_nun_training_64

[7] https://huggingface.co/sentence-transformers/paraphrase-MiniLM-L6-v2, (384 dimensions)
[8] https://www.sbert.net
[9] Python libraries: hdbscan-0.8.28, umap-learn-0.5.3

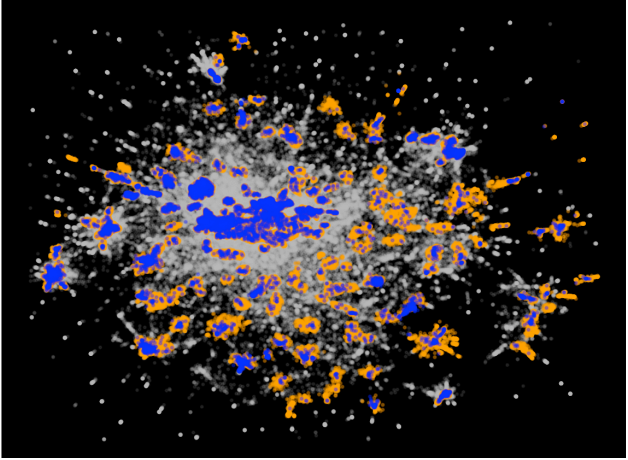 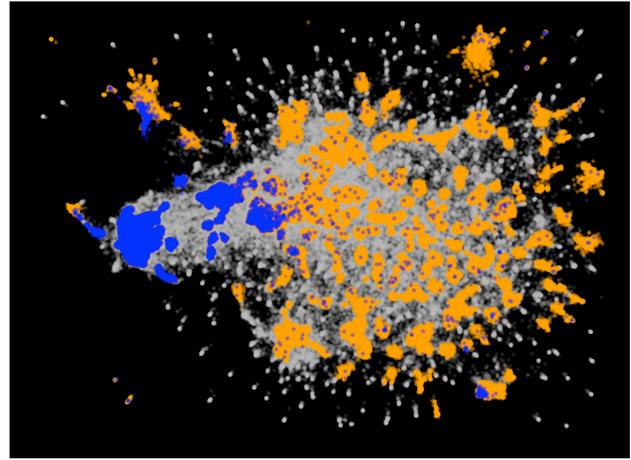

a) Untuned Model       b) Model After Fine-tuning

Figure 2. Comparison of embeddings for untuned (a) and fine-tuned (b) models. Images were developed using UMAP and HDBSCAN with default parameters. Fine-tuned model was developed using *random* sampling with *min* similarity, and a cutoff of 0.8. Areas that could be clustered are colored; orange points are from climate believers, blue points from skeptics.

$$\text{label}(K) = \text{argmax}_k |\{c \in C \mid k \in c\}|$$

Then the purity of cluster **K** may be defined as:

$$\text{purity}(K) = \frac{|\{k \in K : k = \text{label}(K)\}|}{|K|}$$

To determine the parameter ranges for the evaluation step, I manually explored the space to identify points where performance stabilized for each of the techniques introduced above, and then performed more systematic testing in these ranges, as follows: **random**, *cutoff*=[0,.2,.4,.6,.8]; **knn**, *k*=[5,10,15]; **clustered**, *purity*=[.85,.9,.95]. As shown in Figure 3, all approaches increased purity and the number of clusters for skeptic sentences beyond the untuned model, which had none. The purity of believer clusters also increased. The total number of clusters was comparable across the different methods, suggesting that semantic distinctions were generally preserved. The number of skeptic clusters was uniformly low in comparison to believer clusters, reflecting the much smaller number of skeptic sentences.

To select a model for subsequent steps, I normalized each measure to the range [0-1] by dividing by the maximum value within disposition (skeptic / believer) and then summed the values, resulting in an objective function with a range [0-3]. The best model used **random** sampling with a cutoff of zero and the **invert** similarity assignment method. I used this model for all subsequent steps.

**Step 3: Transform beliefs into a belief landscape model and trace belief changes.**
I interpret each cluster in the clustered language model as a belief proposition and use the co-occurrence of these beliefs to construct belief landscape. My approach bears similarity to prior work on belief interdependencies (e.g., Boutyline and Vaisey 2017), but I focus directly on co-occurrence rather than seeking to infer belief interdependencies.

To do this I derived a set of longitudinal belief vectors by sliding a window over each users' tweet history. Each belief vector is a weighted sum of belief statements. I calculated the weight of each statement as an exponential function of time,

$$y_t = \frac{x_t + (1-\alpha)x_{t-1} + (1-\alpha)^2 x_{t-2} + \cdots + (1-\alpha)^t x_0}{1 + (1-\alpha) + (1-\alpha)^2 + \cdots + (1-\alpha)^t}$$

$$\alpha = 1 - e^{-\ln(2)/halflife}$$

where $y_t$ is the calculated belief vector at time *t*, $x_t$ is the observed belief vector, and the *halflife* was adjusted as an evaluation parameter.

To construct the belief landscape, I projected belief vectors into two dimensions with UMAP. To reduce processing time, I sampled 30% of the belief statements for training and used UMAP's supervised learning ability to map the remaining statements into the space. This was the approximate minimum number of samples needed for the spatial organization of the mapped data to mirror that of the training data.

I used visual inspection to set parameters for UMAP, with the following criteria: I sought parameter settings that avoided visual artifacts ("stringiness") while preserving global structure, placed users with the same stance in similar locations, and generated stable results (i.e., approximately isomorphic) over multiple runs. Generally, these goals were met with *nearest_neighbors* in the range [400,600] and *min_dist* in the range [.2,.4]. I evaluated my four hypotheses for nine parameter settings covering this range, with *nearest_neighbors* ∈[400,500,600] × *min_dist*∈ [.2,.3,.4].

Prior to evaluation, I labeled likely bots using the Botometer Lite API (K.-C. Yang et al. 2020), which returns a value from 0-1 indicating the likelihood the user is a bot. The returned data had a central tendency (median = .19, mean = .21) with a noticeable tail beginning around .5. After manually inspecting accounts, I labeled the 544 accounts with a score above .5 as bots (4% of all accounts checked).

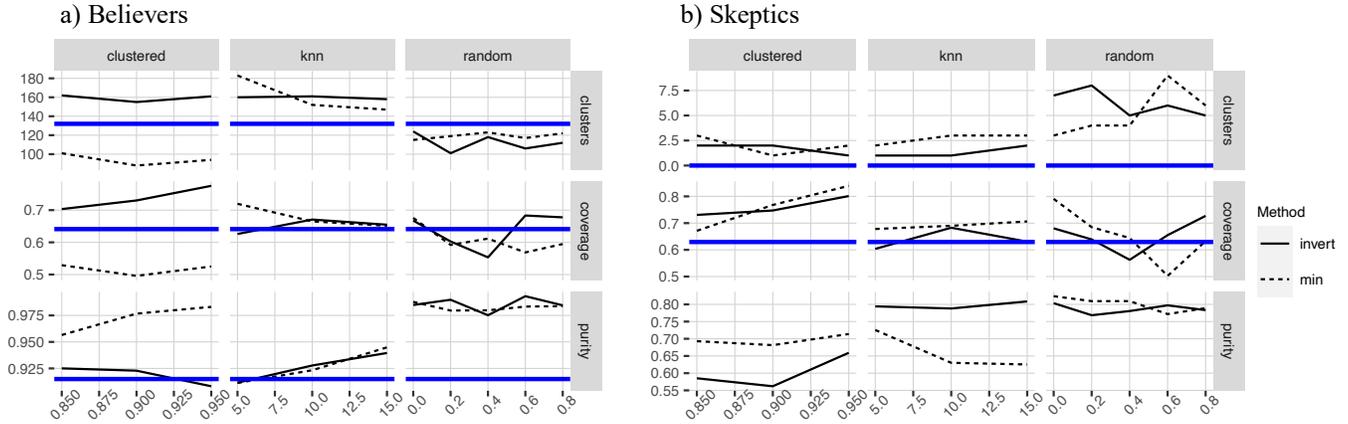

Figure 3. Performance of different training methods for the fine-tuned language model. X-axis values reflect the parameter values used for the different methods, and Y-axis values are counts in the case of clusters, and percentages in the case of coverage and purity. The performance of the untuned model is represented by the horizontal blue lines. Purity could not be calculated for skeptic clusters in the untuned model because no clusters were identified.

**Evaluation**

Hypotheses H1 - H3 are expressed in terms of patterns of movement around attractors. I operationalize attractors as places in the belief landscape that are visited more frequently. To identify attractors, I generated a kernel density estimate from the data developed in Step 3 above, and then identify peaks in the smoothed surface. I used the kernel density function *kde2d* implemented in the MASS library for the R programming language (Venables and Ripley 2002). Mathematically, the kernel density estimate is

$$f(x,y) = \frac{\sum_s \phi((x - x_s)/h_x)\phi((y - y_s)/h_y)}{nh_xh_y}$$

where $n$ is the number of grid points used, $h$ is a bandwidth calculated as

$$h = 4 * 1.06 \min(\hat{\sigma}, R/1.34)n^{-1/5}$$

and $R$ is the inter-quartile range (eq. 5.5, Venables and Ripley 2002, 127). I set the grid size $n=100$ to capture fine-grained detail in the density plot. To identify peaks, I used a second derivative test at all grid points separately along the x and y dimensions and identified points where local maxima overlapped. Specifically, for a given vector of points **x** (here corresponding to the 100 points along a gridline), the set of maxima can be identified as:

$$maxima(\boldsymbol{x}) = \{x|\hat{f}(x) = -2\}$$

$$\hat{f}(\boldsymbol{x}) = \Delta(sgn(\Delta \boldsymbol{x}))$$

where $\Delta$ denotes the empirical slope.

This approach produces a tail of very low amplitude maxima. Analysis revealed a natural break in the data in the region of maxima with a magnitude ~0.2, and so my evaluation only considered all maxima with magnitude > 0.2 (except in the case of H3, discussed below).

I evaluated each hypothesis as follows:

*H1: Attractor stability* – For each user and time window, I map the user to the nearest attractor, and then compute the stability of each user over all time windows:

$$stability = 1 - \frac{\# \ attractors}{\# \ time \ periods}$$

Intuitively, stability is an indicator of how frequently a user jumps between attractors relative to the number of time periods in which they post. I report the distribution of stability for all users who appeared in the data for more than ten time-periods, because there is a great deal of variance among less active users.

*H2: Attractor gradients*—I use logistic regression to determine the how attractor proximity and strength correlate with users' movements to new attractors. In all cases, I include two co-variates: *distance* from the originating attractor and the *strength* of the originating attractor (the magnitude of maxima in the kernel density map).

*H3: Local transitions*—Where users switched attractors, I rank-ordered destination attractors by distance and report the distribution of ranks for the destination. This analysis is sensitive to the number of attractors considered, so I fixed the number of attractors to 20 (the minimum number of attractors with a magnitude > .2 across cases) for each parameter setting to enable direct comparisons.

*H4: Belief homophily*—For each user and time window, I identify the twenty nearest neighbors within a fixed radius

| Window Size (days) | 7 | 14 | 21 | 28 | 35 | 42 |
|---|---|---|---|---|---|---|
| **H1** Stability (mean) | .62 | .66 | .69 | .70 | .72 | .72 |
| **H2** Distance (mean β) | .17 | .23 | .43 | .35 | .34 | .28 |
| Strength (mean β) | -.17 | -.23 | -.23 | -.29 | -.28 | -.37 |
| **H3** Nearby transitions (mean fraction) | .28 | .33 | .39 | .39 | .43 | .44 |
| **H4** Believer homophily (mean) | .98 | .99 | .99 | .99 | .99 | .99 |
| Skeptic homophily (mean) | .62 | .70 | .74 | .78 | .82 | .82 |

Table 1: Summary of statistics over UMAP parameter range. All statistics are mean values; H3 reports the mean percentage of hops within the nearest 5 attractors

(defined as the population average distance users moved between adjacent time windows) and report the percentage of neighbors with the same stance. I evaluated this measure for skeptics and believers separately.

I varied the shape of the exponential decay function by adjusting the *halflife* parameter; for simplicity, I will refer to this as manipulating window size. Choice of window size reflects assumptions about how much data is needed to capture a user's current belief state. Windows that are too small may capture current interests rather than beliefs, and those that are too large may fail to capture dynamics. Because of this I examined the specified hypotheses across a range of window sizes. After this quantitative evaluation, I created a visualization of the landscape and examined representative beliefs across selected attractors.

# Findings

## Quantitative Findings

I examined sliding windows with a half-life from one to six weeks for human users and require users to have been active for 7 days before a first data point is evaluated. Table 1 summarizes results across the parameter range, while Figures 4 & 5 illustrate results with UMAP *nearest_neighbors*=500 and *min_dist* = .4, as these were representative. Findings were consistent across the UMAP parameter range.

*H1: Attractor stability*: There is a central tendency in the data with a mean varying between .58 and .76 for the explored parameter range. The smallest window (7 days) had the lowest value, suggesting it may be too small to identify stable belief states. In summary, I find partial support for H1: most individuals do switch attractors, but this happens infrequently relative to the number of posts they make.

*H2: Attractor gradients*: In all cases, distance from the nearest attractor was positively correlated with the likelihood of transition, strength inversely correlated, and all coefficients were significant at the p<.001 level. On average, the impact of attractor distance is strongest at three weeks, and of attractor strength at four weeks. In summary, I find strong support for H2, but results are sensitive to window size, and the pattern suggests an optimal window size (with respect to H2) between three or four weeks.

*H3: Local transitions*: When transitions are made, there is a preference for attractors that are nearby; across the parameter space, between 25% and 50% of all transitions were less than or equal to rank five, with the percentage increasing with window size. While closer transitions are preferred, a significant proportion of transitions are to more distant attractors. In summary, I find partial support for H3, but note that non-local transitions occur with surprising frequency.

*H4: Belief homophily*: I find strong support for H4. In all cases, people tend to occupy regions of the belief landscape that are close to others with similar stances, though this is more pronounced in the case of climate believers. At all parameter settings, at least 19 out of 20 of a believer's nearest neighbors on the landscape also believe in climate change, regardless of window size. For skeptics, between 42% and 84% of the cases examined had this level of homophily, with a maximum at a window size of 5 weeks. Higher levels of heterogeneity among skeptics are not surprising given the skewed proportions of skeptics and believers. There is bimodality in the case of skeptics (see Figure 4d), indicating that some skeptics "reside" amongst believers on the belief landscape; that is, a small number of skeptics tweet beliefs like those of believers at some points in time. This will be an interesting group of users to explore in future work.

## Qualitative Findings

Figure 5 provides a visualization of the entire landscape, with darker regions and contour lines indicating the location of attractors and their basins. This visualization is based on the analysis with a time-window of five-weeks. I used this window because it seemed to exhibit good performance with respect to the hypotheses discussed above. To help guide intuition about the structure of the landscape, I also randomly sampled 0.1% of the data and plotted these individual points on the landscape, using color to indicate stance.

To interpret attractors, I selected the top five users who had the closest approach to an attractor, and then examined their tweets at that point. Table 2 presents a summary of my examination for the labeled attractors in Figure 5, along with selected tweets (edited to preserve anonymity).

Tweets in the strongest attractor in the landscape (area 1 in Figure 5) are about climate activism in the U.S. They focus on the urgency of climate action and emphasize a moderate liberal agenda (i.e., strong support for Biden as a candidate), with less enthusiasm for the progressive wing of the U.S. Democratic Party. The second strongest attractor (area 2) focuses on the environmental consequences of climate change. These tweets address global impacts covering Asia

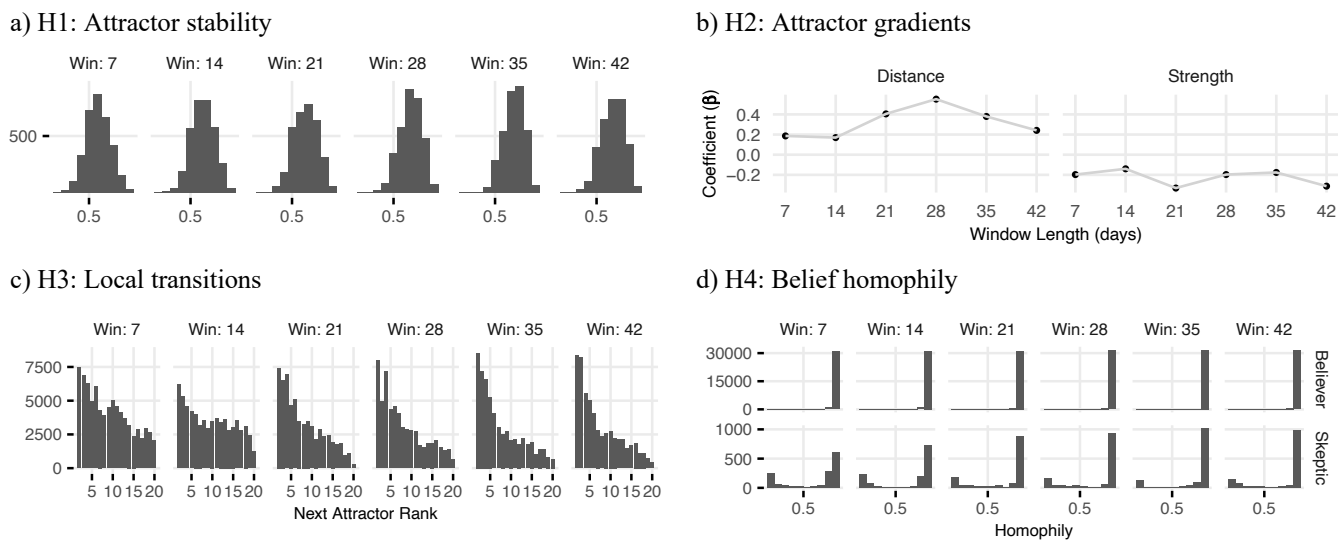

Figure 4. Findings for the four hypotheses at UMAP settings *nn=500, min.dist=.4* a) H1: Most users visit only a small proportion of attractors relative to their level of activity; b) H2: Distance from an attractor correlates with likelihood of moving to a new attractor, and the strength of the attractor is inversely correlated with the likelihood individuals move away from it; c) H3: Individuals who transition between attractors prefer closer attractors; d) H4: Both skeptics and believers are closer to others with the same stance on the belief landscape, though there is slight bimodality in the case of skeptics.

and Europe as well as the U.S. Discussions about the wildfires in the U.S. are also prevalent here.

Beliefs in more peripheral attractors are somewhat outside of the Twitter mainstream. Area 3 (the ninth ranked attractor) dwells on climate denialism, focusing on denialists themselves or those who call out denialism-based government inaction. Areas 4 and 5 on the map are predominantly skeptics' regions. Area 4 is the third strongest attractor in the landscape, and the bulk of skeptics' activity occurs here. Tweets in this region tend to offer the opinion that the climate may be changing and that humans may have a role, but there is no crisis and efforts to address the problem are an unnecessary distraction. Area 5 (one of the weaker attractors on the landscape) takes these claims further, implying or efforts to address climate change are motivated by greed and conspiracies. Tweets in this area often point to the UN as central conspirator, and Canada as a victim.

In summary, the map reflects many nuanced pools of belief related to different aspects of climate, and articulated beliefs are spatially arranged in a manner that corresponds to perceivable semantic relationships.

## Discussion

My findings indicate that the BLF generates an analysis consistent with a range of theory-based expectations about belief dynamics. While a promising avenue that could address a gap in research on belief dynamics, it remains exploratory. Below, I discuss both limitations and implications.

The size of the parameter space is a significant limitation of the method. In the language model alone, there are many possibilities for deriving a semantic space, and these have unexamined consequences for the structure of the belief landscape. How time windows are computed also impacts results and requires further exploration. The parameter settings I have examined are data-driven and lack theoretical motivation because strong theoretical guidance is not available. For instance, how the embedding space of deep language model should be spatially organized to approximate human understanding is central to the active area of Interpretable AI (Doshi-Velez and Kim 2017).

A related limitation stems from my reliance on UMAP to establish the structure of the belief landscape. McInnes et al. (2020) warn against over-interpretation of the spatial arrangement of data in low-dimensionality projections. Quoting from online materials[10], "UMAP does not completely preserve density... [and] can also create false tears in clusters, resulting in a finer clustering than is necessarily present in the data." My evaluations depend on the spatial structure of the landscape, and while not anomalous for the explored region of the parameter space, choosing this region of the parameter space was not theoretically motivated.

Another challenge is the lack of an objective function or gold data for evaluating results; absent theory-driven parameters, conformity with gold data could increase confidence in the method. We might seek to tune the method with the goal of predicting individuals' movements, but this ethically complicated, which I consider below.

---

[10]https://web.archive.org/web/20221201223855/https://umap-learn.readthedocs.io/en/latest/clustering.html

| Area Summary | Prototypical Tweets (two different users) | |
|---|---|---|
| **Area 1 – U.S. politics oriented, mainstream liberal climate beliefs** | The reason Biden's climate proposal are increasingly ambitious is because of public demand! | Biden, unlike Bernie, doesn't suggest we should ban the internal combustion engine! His climate plan is realistic, actionable, and ensures the US will achieve 100% clean energy and net-zero emission by 2050. |
| **Area 2 – Environmental impacts of climate change; wildfires** | It should be obvious to everyone that these wildfires are driven by fossil fuels – fossil fuels cause climate change, climate change causes more intense wildfires. | Humanity's breadbasket are the great river deltas of the world, supplying many stable crops. Many of these areas are exceedingly vulnerable to changing sea levels. |
| **Area 3 – Focus on climate denialism and gov't inaction.** | Meteorologist in Hong Kong 'ashamed' of Hong Kong Climate Policy | First, Joni Ernst denied climate science, and now she's attacking doctors with Covid conspiracy theories! |
| **Area 4 – Denial of a climate "crisis"** | @Frodo655 The point we both agree on, is that we are clearly NOT in a climate crisis. We might differ on details, but that's not important here. | People are terrified about things they don't need to be. The climate changes, but will higher taxes stop bad weather? Nope! |
| **Area 5 – Canada focused, anti-UN, conspiracy laden skepticism** | UN is changing its narrative on Covid19 because lockdowns have halted the transfer of wealth, setting back gains made by the climate change scam. | Greta Thunberg wants to turn Canada into a UN-inspired theme park. Here's to progress! |

Table 2: Description of labeled areas in Figure 5, along with prototypical tweets from each region

There are also deeper questions about whether the method in fact reveals something about belief. The analysis is a type of dynamic, correlated topic analysis over a subset of online posts. Whether such linguistic activity is a good proxy for belief is unclear. Given that this likely follows salient world events (e.g., press releases, disasters, speeches, revelations), it is possible we are observing general collective conversational trends rather than the structure of belief. Future work to allay such concerns should seek to ground this data-driven analysis in richer qualitative studies.

My analysis is also platform-specific and neglects non-English statements. The method does not depend on these limitations; other language models are available, and the approach is viable for any text corpus. Nonetheless, the landscape analysis presented here is biased by these selection criteria. Further development of the method and examinations of other sources could afford more powerful insights.

Limitations aside, my results offer a first view of belief dynamics in a system where coherent belief states are not defined by two ends of a normative ideological spectrum (e.g., conservative vs. liberal, believer vs. skeptic). The analysis raises important questions about the structure of the space, and how populations are likely to traverse it. Opening the door to a multiplicity of coherent belief-states immediately raises questions about how many there are likely to be, how frequently they are created, and how they change. An important next step will be to extend the analysis to a broader range of topics to examine how the number of attractors scales with the breadth of conversation examined.

My initial data also suggests that the belief-landscape itself is reasonably well correlated, and that people tend to follow a gradient towards more coherent states. This is intuitively plausible but raises many questions about behavior and psychology. Do some people move amongst belief states with greater stochasticity? Can belief change happen suddenly, cascading through an individual's complete set of beliefs, or is it more gradual? Further study is necessary, and my approach helps to pave the way to such studies.

## Broader Impacts and Ethical Considerations

An important goal of this work is to advance ongoing research about how media, misinformation, and design might influence belief dynamics. I have demonstrated the method using a narrow set of data from a single platform, but it is relatively straightforward to adapt the approach to other text-based forms of social media (e.g., Reddit, Facebook) and to examine the fluctuation of dynamics in relation to different circumstances. This is one of the more impactful applications of the Belief Landscape Framework. There are many cross-disciplinary discussions about the role media and misinformation play in polarization and the sustenance (or even amplification of) exclusionary belief structures but untangling the causal forces at play is complex.

The BLF could shed new light on these causal interactions by revealing the impact of numerous factors (e.g., a single fake news article or a coordinated media campaign) on belief dynamics. Methods such as surveys, sentiment analysis, or stance detection offer a similar kind of insight, but with less precision. A refined version of the BLF would be unprecedented in the granularity of insight it affords and speed with which it can detect changes in belief dynamics.

The BLF raises many important ethical questions. The method might be improved to make accurate predictions about how different forms of media exposure influence, for instance, the dynamics of anti-vax or racist beliefs. This could be used to develop more effective interventions to move beliefs in either direction. There is already concern about how algorithms intended to shape engagement with online content may have unintentional side-effects on belief and social dynamics. The technology envisioned here could be used to manipulate belief intentionally, and it is therefore urgent that, as a society, we begin to develop regulatory and legal frameworks to grapple with this. It is critical that the research community has an active role in such discussions.

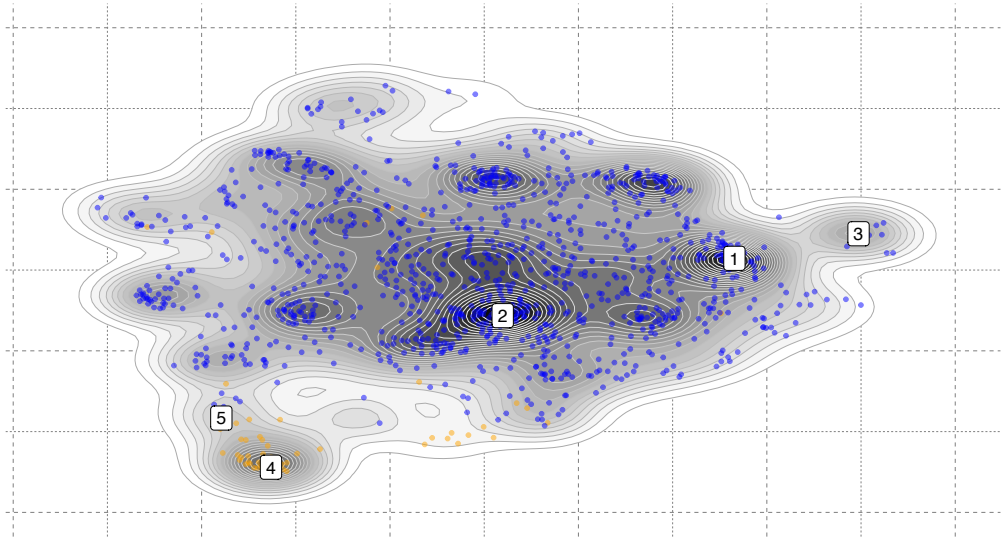

Figure 5. Visualization of the belief landscape. Blue dots reflect the position of climate believers, orange of climate skeptics. Each point is a specific user at a specific point in time. 0.1% of all data points are shown. Labeled areas are selected attractors, described in Table 2.

It is also important to consider individual privacy and autonomy. In this paper, I have sought to protect individual's anonymity by editing quoted tweets, but more advanced versions of this work will need to address other challenges. For instance, a method like this could be used to make predictions about specific individuals, e.g., whether someone is likely to become a skeptic, fall for misinformation, or join an extremist network. It is therefore important to consider the ethical implications of predictive applications.

Such concerns should not prevent further pursuit of these methods, because they can help us better understand how media and technology influence belief dynamics. Rather, pursuing this inquiry in dialog with a research community that values the social good is one of the better ways to minimize the likelihood they will be used to do harm.

## Conclusion

Beliefs and how they change are central to concerns about misinformation, polarization, extremism, and toxicity. Despite this centrality, researchers have not yet developed tools that would enable powerful inferences about how different mediating influences shape belief dynamics. The approach introduced here lays the groundwork for such tools and reveals novel, if tentative, observations about belief dynamics. Used with the appropriate care and ethical awareness, these methods could dramatically enhance our understanding of how the broader digital media ecosystem impacts the belief dynamics of our technologically mediated society.

## Acknowledgements

This work would not have been possible without the input and support of many collaborators, including Jeff Hemsley, Dipto Das, Akit Kumar, and Lizhen Liang. I am also grateful for insightful and detailed feedback from the anonymous reviewers, which helped improve the quality of this work.